{}
{}
\documentclass[english,11pt,aps,prd,superscriptaddress,preprintnumbers,nofootinbib,preprint,titlepage]{revtex4-1} 

\usepackage{multirow}
\usepackage{amsmath} 
\usepackage{amssymb} 
\usepackage{amsfonts} 
\usepackage{graphicx} 
\usepackage{xcolor}
\usepackage[labelformat=simple]{subcaption}

\usepackage[utf8]{inputenc}
\usepackage[english]{babel}

\usepackage{cleveref}
\crefformat{section}{Section~#2#1#3}
\crefformat{figure}{Fig.~#2#1#3}
\crefformat{table}{Table~#2#1#3}
\crefformat{equation}{Eq.~(#2#1#3)}
\crefrangeformat{equation}{Eqs.~(#3#1#4) to~(#5#2#6)}
\crefmultiformat{equation}{Eqs.~(#2#1#3)}{ and~(#2#1#3)}{, (#2#1#3)}{ and~(#2#1#3)}
\crefrangemultiformat{equation}{Eqs.~(#2#1#3)}{ and~(#2#1#3)}{, (#2#1#3)}{ and~(#2#1#3)}
\crefname{appsec}{Appendix}{Appendices}

\newcommand{\be}{\begin{equation}}
\newcommand{\ee}{\end{equation}}

\allowdisplaybreaks

\begin{document} 

\title{Internal Reduction method for computing Feynman Integrals} 

\preprint{TUM-HEP-1232/19} 
\def\NCSR{Institute of Nuclear and Particle Physics, NCSR Demokritos, Agia Paraskevi, 15310, Greece}
\def\TUM{Physik-Department T31, Technische Universität München, James-Franck-Straße 1, D-85748 Garching, Germany} 

\author{Costas G. Papadopoulos} 
\email[Electronic address: ]{costas.papadopoulos@cern.ch} 
\affiliation{\NCSR} 

\author{Christopher Wever} 
\email[Electronic address: ]{christopher.wever@tum.de} 
\affiliation{\TUM} 

\begin{abstract} 
A new approach to compute Feynman Integrals is presented. 
It relies on an integral representation of 
a given Feynman Integral in terms of simpler ones. 
Using this approach, 
we present, for the first time, results for a certain family of non-planar five-point two-loop Master Integrals with one external off-shell particle, relevant for instance for $H+2$ jets production at the LHC, 
in both Euclidean and physical kinematical regions.
\end{abstract} 

\maketitle
\pagenumbering{gobble}

\setcounter{page}{1}
\flushbottom

\pagenumbering{arabic}

\section{Introduction } 
\label{sIntro}

Precise theoretical predictions are nowadays a vital element of discoveries in Physics, from gravitational wave astronomy~\cite{Abbott:2016blz} to high-energy physics~\cite{Aad:2012tfa,Chatrchyan:2012xdj}. The coming LHC Run 3 and the High Luminosity LHC Run scheduled after it, require the most precise theoretical predictions in order to fully exploit the machine’s potential~\cite{Bendavid:2018nar}. In the future, the FCC (Future Circular Collider) project will also further boost the demands in the direction of precision~\cite{Mangano:2017tke}.   

To be precise, next-to-next-to-leading order (NNLO) accuracy is the next level needed for physics analysis at the LHC, for the vast majority of QCD dominated scattering processes (see~\cite{Heinrich:2017una} and references therein). Mixed QCD-electroweak effects are also important for processes and observables involving gauge bosons productions~\cite{Dittmaier:2015rxo,Bonciani:2016ypc,Delto:2019ewv}, whereas when the NNLO corrections turn out to be large, even higher loop calculations (N$^3$LO) are required~\cite{Anastasiou:2015ema}.

Over the last years, NNLO QCD corrections for most of the $2\to 2$ processes, including two-jet, top-pair and gauge bosons production, have been completed and already used in phenomenological and experimental studies~\cite{Azzi:2019yne}. This is the result of an intense theoretical work, which can be schematically classified in two frontiers: the frontier of NNLO radiative corrections and the frontier of two-loop amplitude computation. Building on the progress in these two frontiers, first results on NNLO QCD corrections for $2\to 3$ processes start to emerge~\cite{Badger:2017jhb,Abreu:2017hqn,Abreu:2019odu,Hartanto:2019uvl}. 

Two-loop amplitude computations require the reduction of the scattering matrix element in terms of basis integrals, usually referred to as Master Integrals (MI). Traditional reduction techniques based on integration-by-part identities~\cite{Chetyrkin:1981qh,Tkachov:1981wb,Laporta:2001dd} (IBP) are now more and more replaced by integrand reduction methods, following the one-loop paradigm~\cite{Ossola:2006us}. Results for five-point two-loop amplitudes, relevant for three-jet/photon, $V,H+2$ jets production have been recently presented~\cite{Hartanto:2019uvl}. A remarkable contradistinction with the NLO case, is that the basis of Master Integrals at two loops is still incomplete. 

Multi-loop Master Integrals have been investigated for many years now. The most successful method to obtain analytic expressions and accurate numerical estimates of multi-scale multi-loop Feynman Integrals is, for the time being, the differential equations approach~\cite{Kotikov:1990kg,Kotikov:1991pm,Bern:1992em,Remiddi:1997ny,Gehrmann:1999as}. With the introduction of the canonical form of the differential equations~\cite{Henn:2013pwa}, a major step towards the understanding of the mathematical structure of Feynman Integrals and subsequently of the scattering amplitudes has been achieved. The complexity of two-loop Feynman Integrals is determined by the number of internal massive propagators and the number of external particles, i.e. the total number of independent "kinematical" scales involved. Feynman Integrals with a relatively small number of scales satisfy canonical differential equations and can be expressed in terms of multiple (or Goncharov) polylogarithms~\cite{Goncharov:1998kja,Remiddi:1999ew,Goncharov:2001iea}, a class of functions that have been well understood by now.   
Moreover, in the last couple of years, new mathematical structures~\cite{Adams:2015gva,Bonciani:2016qxi,Ablinger:2017bjx,Bourjaily:2017bsb,Broedel:2018iwv} (elliptic polylogarithms) have been studied in order to obtain analytic insight of more complicated Feynman Integrals.
With a complete basis of two-loop Master Integrals, it is hoped that an automation of NNLO calculations for arbitrary scattering processes can be achieved in the near future. 

\begin{figure}[t!]
\centering
\includegraphics[width=0.20 \linewidth]{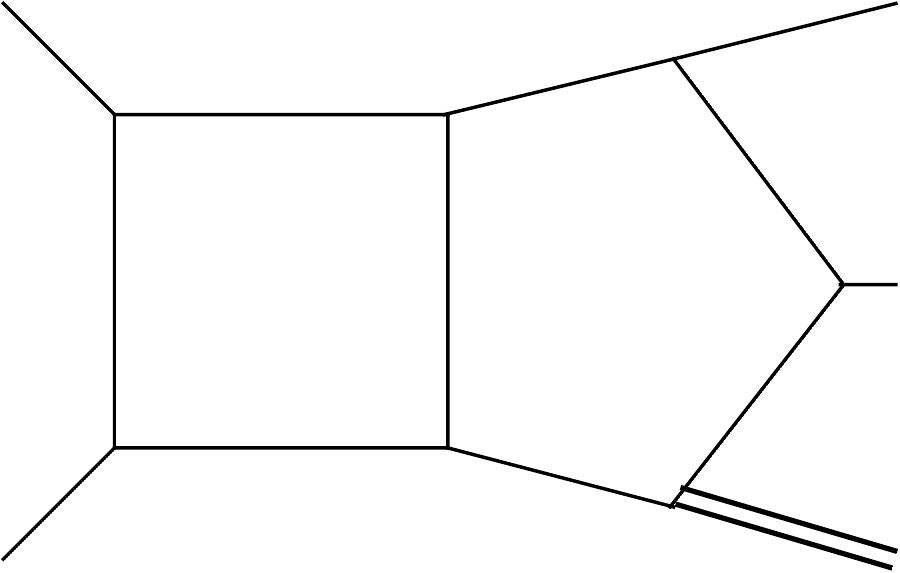} \hspace{0.4 cm}
\includegraphics[width=0.20 \linewidth]{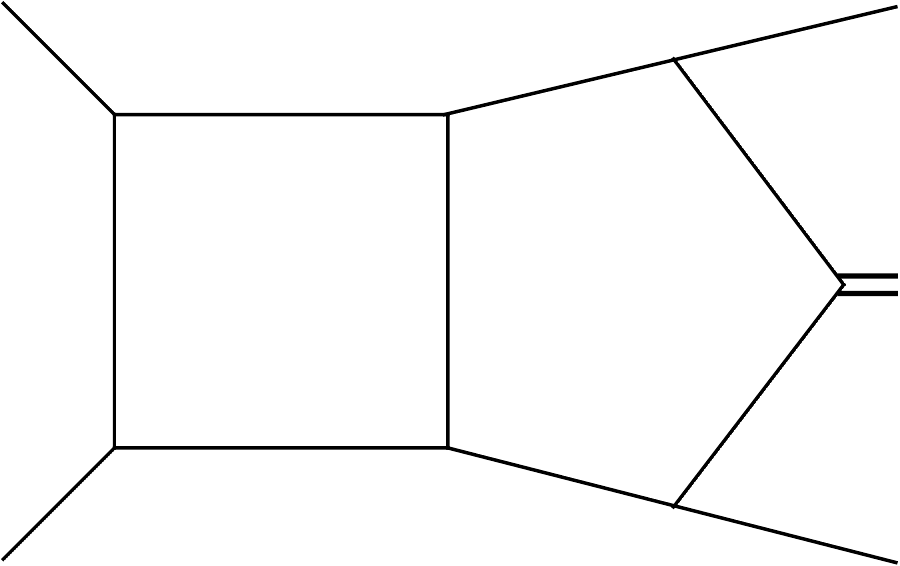} \hspace{0.4 cm}
\includegraphics[width=0.20 \linewidth]{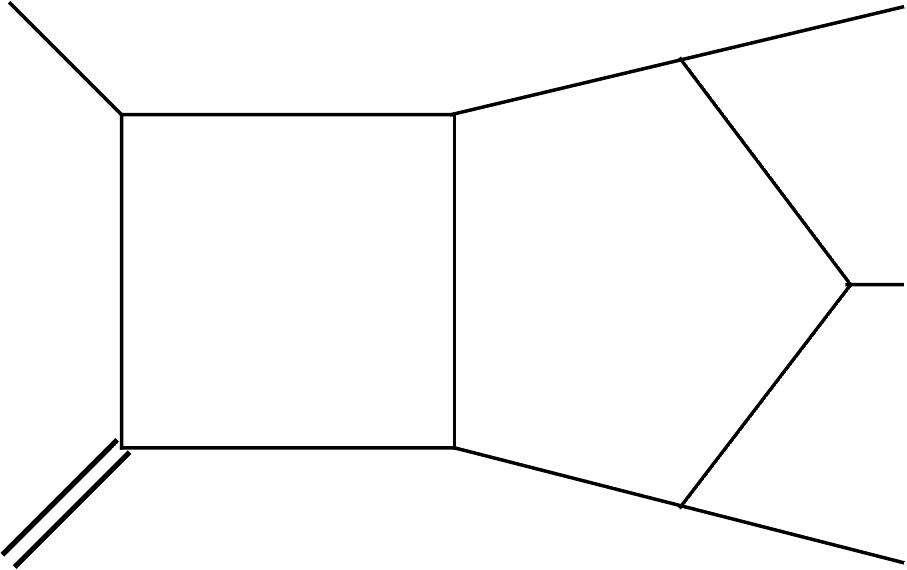}
  \caption{The three planar pentabox families: $P_1$ (left), $P_2$ (middle) and $P_3$ (right) with one external massive leg.}
  \label{fig:param-P}
\end{figure}

\begin{figure}[t!]
\centering
\includegraphics[width=0.20 \linewidth]{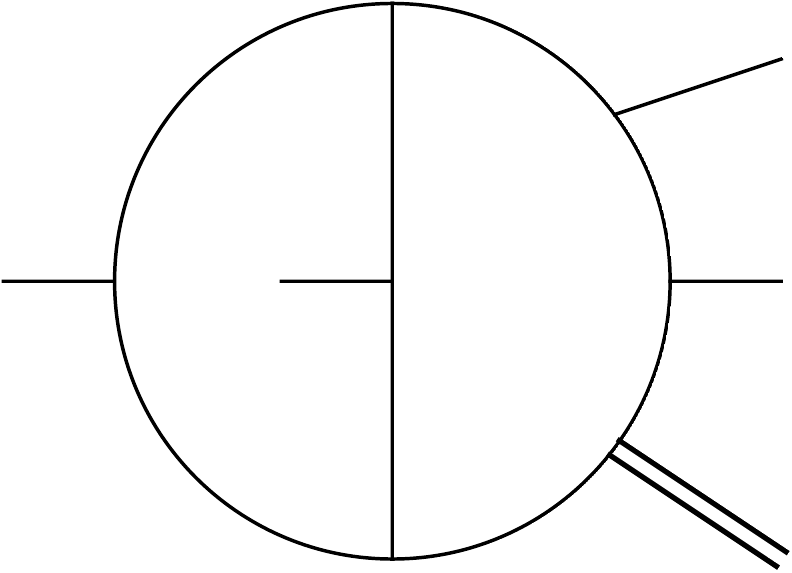} \hspace{0.6 cm}
\includegraphics[width=0.20 \linewidth]{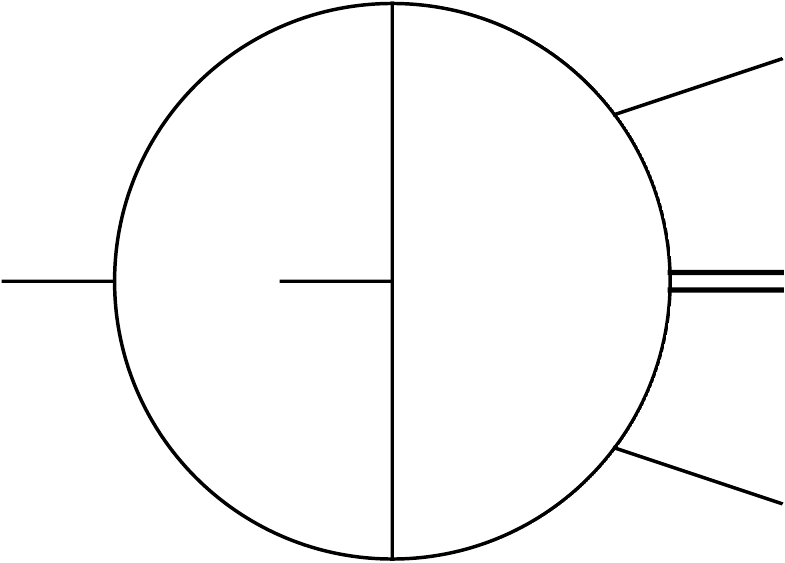} \hspace{0.6 cm}
\includegraphics[width=0.20 \linewidth]{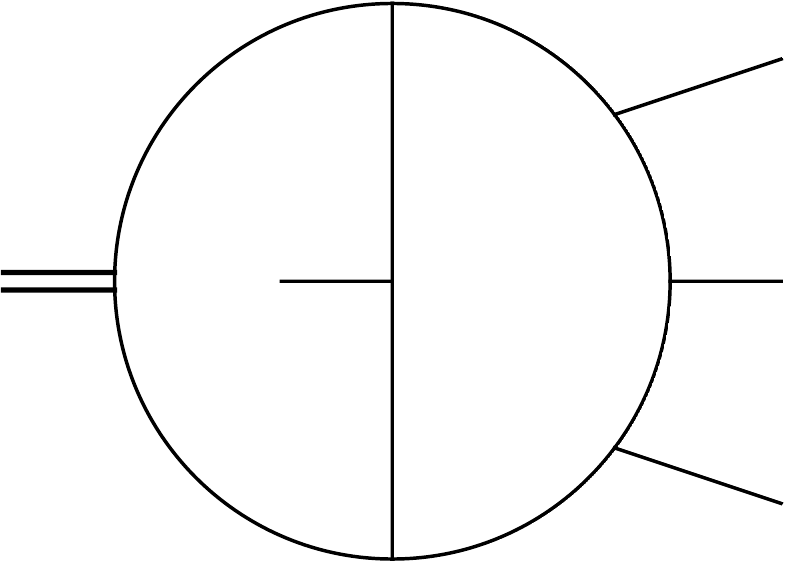} \\[12pt]
\includegraphics[width=0.20 \linewidth]{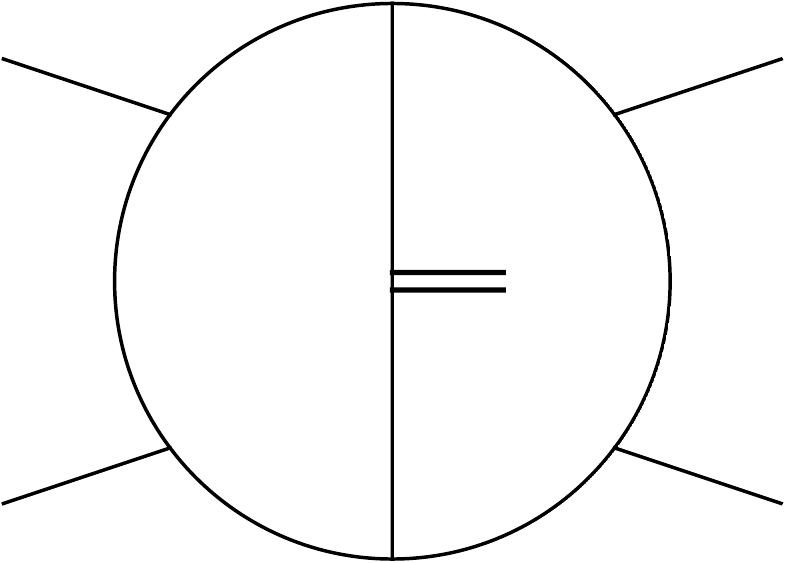} \hspace{0.6 cm}
\includegraphics[width=0.20 \linewidth]{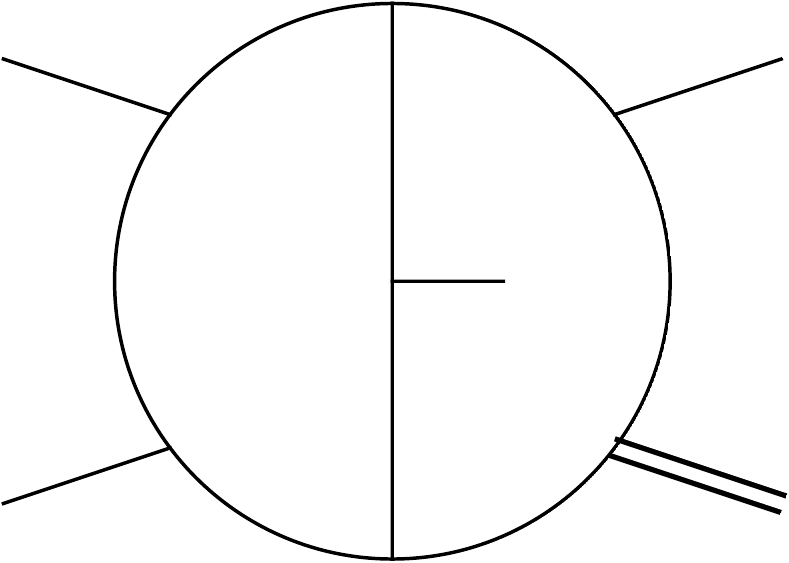}
  \caption{The five non-planar families with one external massive leg: $N_1$ (top left), $N_2$ (top middle), $N_2$ (top right), $N_4$ (bottom left), $N_5$ (bottom right)}
  \label{fig:param-N}
\end{figure}

Five-point two-loop Master Integrals determine the current frontier of this endeavour. The computation of all planar and non-planar five-point two-loop Master Integrals with massless internal propagators and light-like external momenta, has been recently completed~\cite{Gehrmann:2015bfy,Chicherin:2018old}. The next step on this path of computing the five-point two-loop Master Integrals would be those with one of the external legs being off-shell. The planar and nonplanar topologies corresponding to these Master Integrals are shown in Fig.~\ref{fig:param-P} and Fig.~\ref{fig:param-N} respectively. Based on a simplified differential equations approach, we have also computed and expressed in terms of Goncharov polylogarithms, all Master Integrals for the first non-trivial planar family of five-point two-loop Master Integrals with massless internal propagators and one external particle carrying a space- or time-like momentum, $P_1$ in Fig.~\ref{fig:param-P},
as well as the full set of planar five-point two-loop massless Master Integrals with light-like external momenta~\cite{Papadopoulos:2015jft}. In this paper we introduce a new idea, that we call "Internal Reduction", and present for the first time results for the 
non-planar family of five-point two-loop Master Integrals with massless internal propagators and one external particle carrying a space- or time-like momentum, $N_1$ 
in Fig.~\ref{fig:param-N}.

In section \ref{sIntRed}, we briefly introduce the idea and present some key examples of its applicability. Section \ref{sPenta} gives a detailed presentation of the computation of the non-planar five-point two-loop Master Integrals, based on the Internal Reduction approach. Finally in section \ref{sResults} we present our results, and whenever possible, a comparison with purely numerical approaches based on sector decomposition. Finally in section \ref{sConc} we discuss the future applications of the Internal Reduction approach, with emphasis on the completion of the computation of all planar and non-planar five-point two-loop Master Integrals considered.    

\section{Internal Reduction} 
\label{sIntRed}

As two-loop Feynman Integrals with many scales (many external legs and/or internal masses) become more and more difficult to be expressed in analytic form, it is very welcome to investigate the possibility to compute them in terms of simpler ones, for which analytic results are available.   
In this respect, a Feynman parameter is introduced, in order to appropriately combine internal 
propagators of a multi-loop Feynman Integral. 
The simplest case is to combine two neighbouring propagators with the same loop momentum 
\begin{equation}
\frac{1}{{ \cdots \left[ {{{\left( {k + {p_1}} \right)}^2} - m_1^2} \right]\left[ {{{\left( {k + {p_2}} \right)}^2} - m_2^2} \right] \cdots }} = \int\limits_0^1 {dx\frac{1}{{{ \cdots{\left[ {{{\left( {k + q} \right)}^2} - {M^2}} \right]}^2\cdots}}}}
\label{xf}
\end{equation}
with
$q = x{p_1} + \left( {1 - x} \right){p_2}$
and
${M^2} = xm_1^2 + \left( {1 - x} \right)m_2^2 - x\left( {1 - x} \right){\left( {{p_1} - {p_2}} \right)^2}$. 
By appropriately choosing the propagators to be combined, 
the resulting Feynman Integral that appears under the integral of the introduced Feynman parameter $x$, i.e. the integrand, which contains one internal line less corresponds then either to a simpler topology with the same number of external legs or to a Feynman Integral with less external lines.
The task then is straightforward: using known results for the integrand with one fewer internal line, the original Feynman Integral can be computed as an one-dimensional integration, done preferably analytically or in the worst case numerically. In the latter case, one expects that numerical integration shall be much more efficient, in comparison with more traditional approaches based on
sector decomposition or Mellin-Barnes representation. We will call this process {\it Internal Reduction}, since the integrand contains the squared propagator on the r.h.s. in Eq.~\eqref{xf} and will typically itself be reducible to Master Integrals of the topology of which the integrand belongs to. Since this topology would contain one propagator less, its IBP reduction would be typically much easier than the reduction of integrals in the sector of the the original Feynman integral, that contains instead the two propagators on the l.h.s. of Eq.~\eqref{xf}.


It is usually preferable to combine neighbouring propagators separated by an external leg carrying a light-like momentum, since in that case the new internal propagator is still massless. In several case, the reduced integral can be easily expressed in an analytic form using standard techniques, as for instance IBP identities, differential equations, etc. To illustrate the way Eq.~(\ref{xf}) works in that case, a rather trivial example, shown graphically below
\begin{equation}
\qquad
\begin{gathered}
\vcenter{\hbox{\includegraphics[scale=.35]{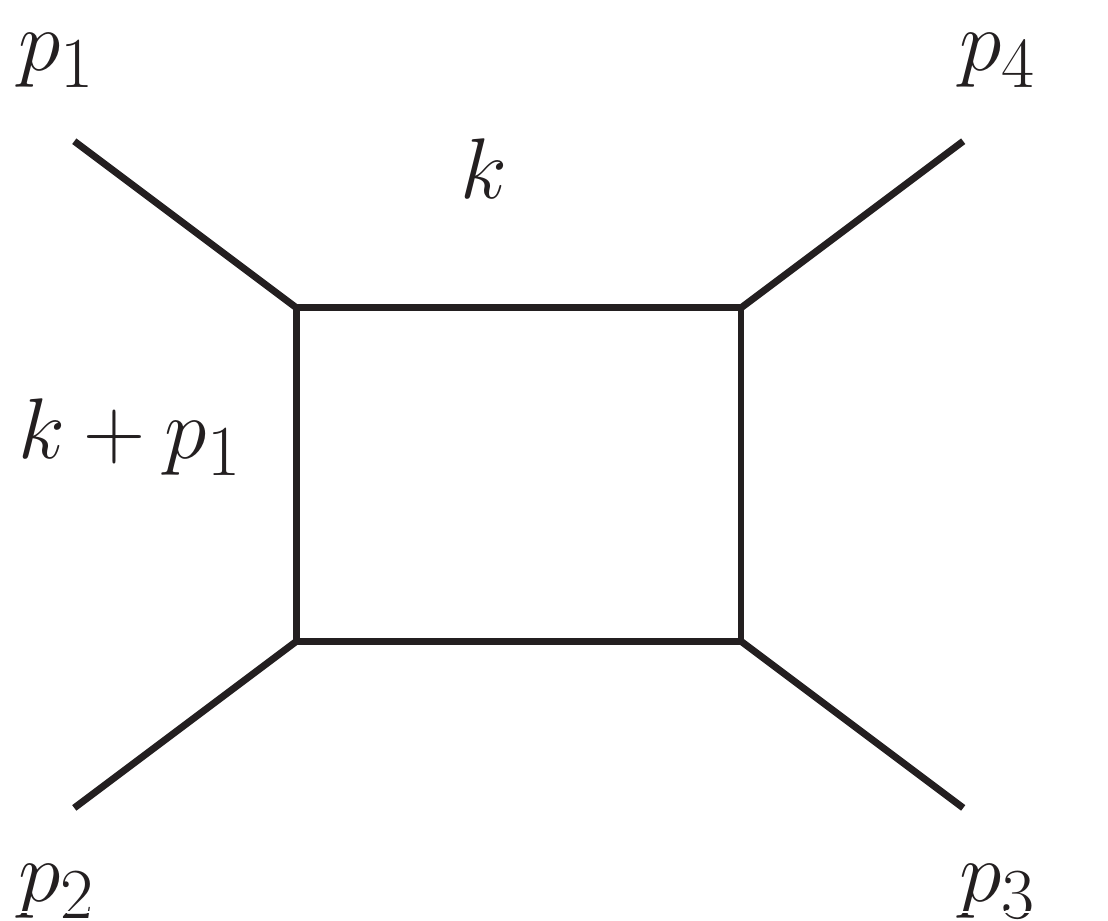}}}\qquad=\qquad\int_{0}^{1} dx\qquad  \vcenter{\hbox{\includegraphics[scale=.35]{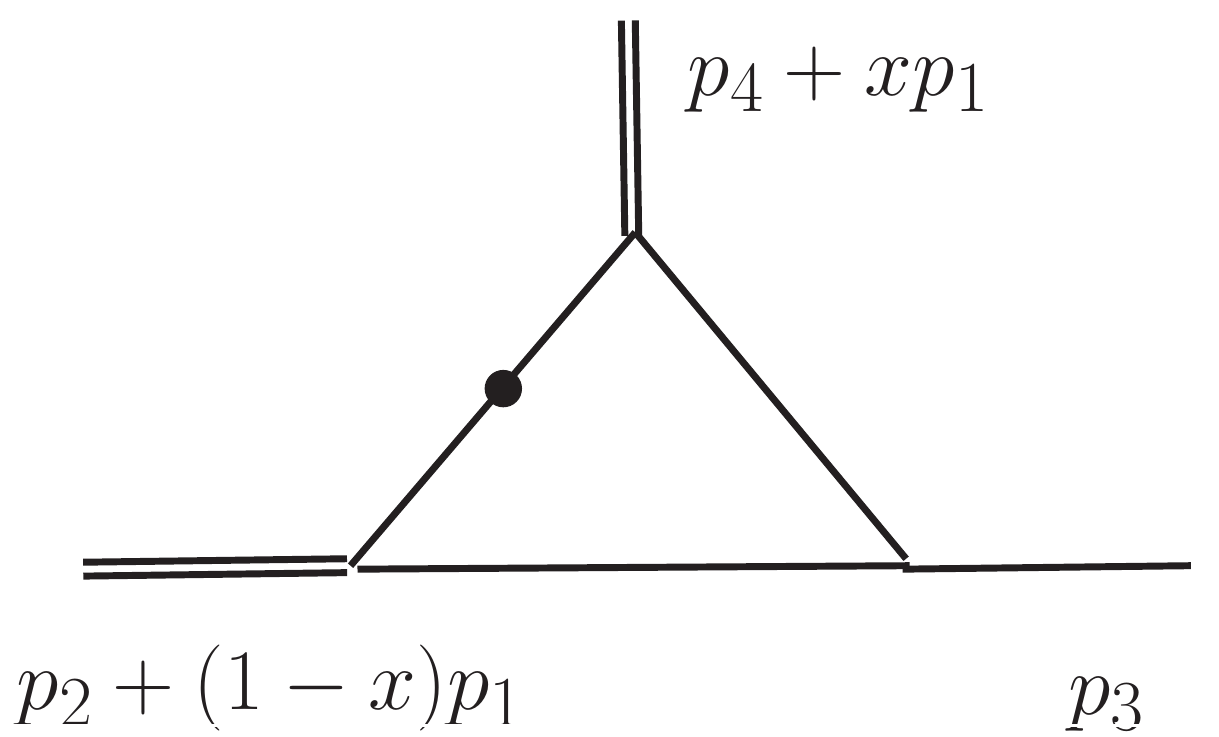}}}
\end{gathered}
\label{1l4pto1l2p}
\end{equation}
has been considered in detail in \cite{Papadopoulos:2018hbt}.
If the two neighbouring propagators are separated by an external leg carrying a time-like or a space-like momentum, the reduced graph will exhibit a non-zero internal mass. To illustrate how the method works in that case, we consider the following example,
\begin{equation}
\qquad
\begin{gathered}
\vcenter{\hbox{\includegraphics[scale=.35]{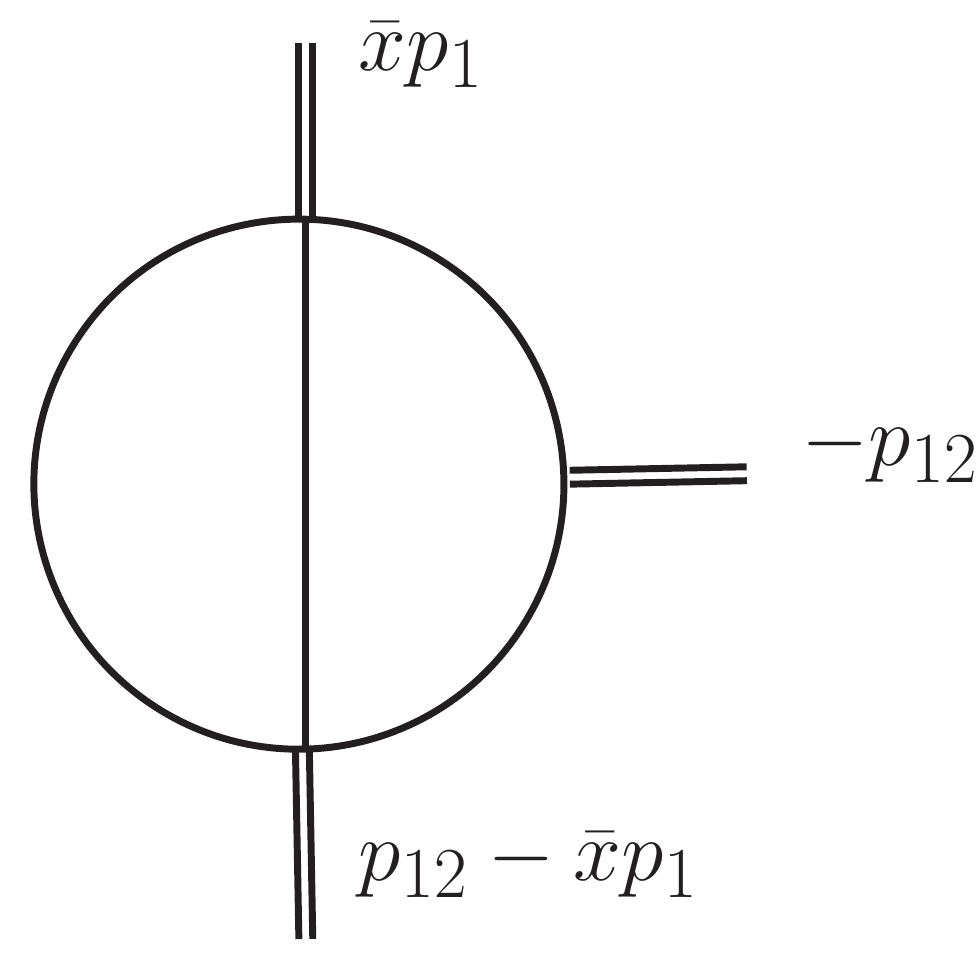}}}\qquad=\qquad\int_{0}^{1} dx\qquad  \vcenter{\hbox{\includegraphics[scale=.35]{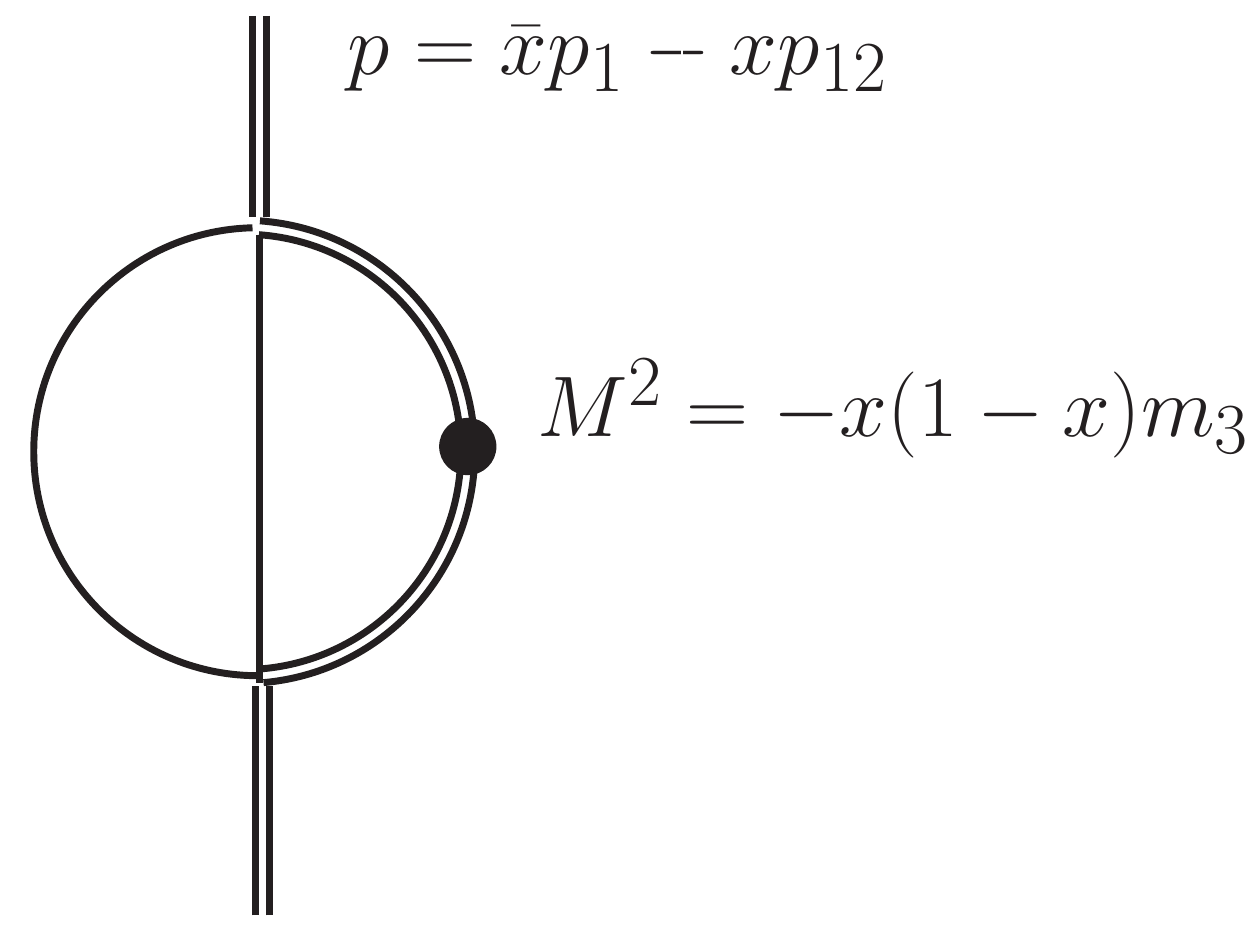}}}
\end{gathered}
\label{2l3pto2l2pm}
\end{equation}
In Eq.~(\ref{2l3pto2l2pm}), $p_1^2=m_1$, $p_2^2=0$, $p_{12}^2=m_3$ and the internal massive propagator is given by
$P^{-1}={\left( {{k_2} - x{p_{12}}} \right)^2} + x(1-x) m_3$. 
Using IBP identities we can straightforwardly express the integrand appearing in the right-hand side of the Eq.~(\ref{2l3pto2l2pm}) in terms of known Master Integrals~\cite{Remiddi:2016gno},
\qquad
\begin{align}
{\cal I}\left(0,1,2,1,0\right)&=
-\frac{(-7+2 d)(m^2-s)}{(d-3)(3 d-10)}\;\;\;{\cal I}\left(0,2,1,2,0\right)\nonumber\\
&-\frac{(-18+5 d)m^2+(3 d-10) s}{(d-3)(3d-10)}\;\;\;{\cal I}\left(0,2,2,1,0\right).
\end{align}
\label{kite-reduced}

Now using the analytic results of~\cite{Remiddi:2016gno} for the above Master Integrals, the integrand in Eq.~(\ref{2l3pto2l2pm})
is, up to an overall factor of $(M^2)^{-2\epsilon}$, with $d=4-2\epsilon$, a function of the dimensionless variable $u=p^2/M^2=\frac{(\bar{x}-x)(m_1 \bar{x}-x m_3)}{-x(1-x)m_3}$, namely
\begin{align}
(M^2)^{-2\epsilon}\left(\frac{1}{2 \epsilon ^2}+\frac{1}{2\epsilon }+\frac{\pi ^2-2}{4}+\frac{(u-1) {\cal G}(1;u)}{u}-{\cal G}(0,1;u)+\mathcal{O}(\epsilon)\right).
\end{align}
Using the known properties of multiple polylogarithms ${\cal G}$, the integrand can be expressed 
in terms of multiple polylogarithms with argument $x$ and letters that depend on $m_1,m_3$ and $\bar{x}$.
The integrand in the right-hand side of Eq.~(\ref{2l3pto2l2pm}) is then easily integrable in the interval $x\in[0,1]$ and a straightforward integration results in the l.h.s. of Eq.~\eqref{2l3pto2l2pm},
\begin{gather}
\frac{1}{2\epsilon^2}+\frac{\frac{5}{2}-\log (-m_3)}{\epsilon}+
\frac{19}{2}-\frac{\pi^2}{12}-{\cal G}\left(0,1,\bar{x}\right)
-{\cal G}\left(0,\frac{m_3}{m_1},\bar{x}\right)
+\frac{2 m_1 (\bar{x}-1) {\cal G}(1,0,\bar{x})}{m_1-m_3}\nonumber\\
+\frac{2 (m_3-m_1 \bar{x}) {\cal G}\left(\frac{m_3}{m_1},0,\bar{x}\right)}{m_1-m_3} 
-\frac{ m_1 (\bar{x}-1) {\cal G}(1,\bar{x}) \log \left(\frac{m_3}{m_1}\right)}{m_1-m_3}
-\frac{ (m_3-m_1 \bar{x}) \log \left(\frac{m_3}{m_1}\right) {\cal G}\left(\frac{m_3}{m_1},\bar{x}\right)}{m_1-m_3} \nonumber\\
-5\log (-m_3)+\log (-m_3)^2,
\end{gather}
that, after replacing $\bar{x} \to x$ above, agrees with the known results of ref.~\cite{Papadopoulos:2014lla}. In fact, although results are shown above up to  $\mathcal{O}(\epsilon)$ for the sake of brevity, we have checked the result up to order
$\mathcal{O}(\epsilon^2)$, namely up to transcendental weight four.  
Notice that, in this case, despite the fact that the Internal Reduction produces an integral with an internal mass, solutions of the sunrise graph, obtained from the differential equations approach, do allow for
a straightforward analytic evaluation of the two-loop triangle, using Eq.~(\ref{2l3pto2l2pm}).

\section{Application to Pentabox Integrals} 
\label{sPenta}

\begin{figure}[t!]
\centering
\includegraphics[width=0.3 \linewidth]{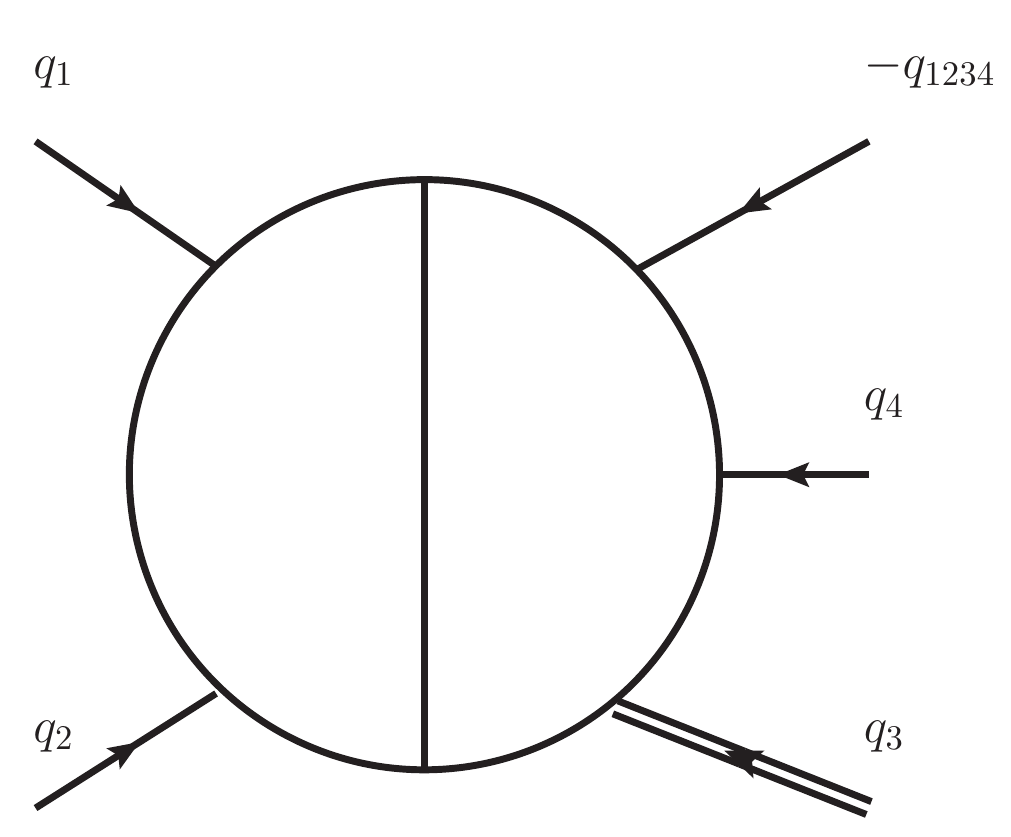}
\caption{The two-loop planar pentabox family $P_1$. All external momenta are incoming.}
\label{fig::diagp1}
\end{figure}

\begin{figure}[t!]
\centering
\includegraphics[width=0.3 \linewidth]{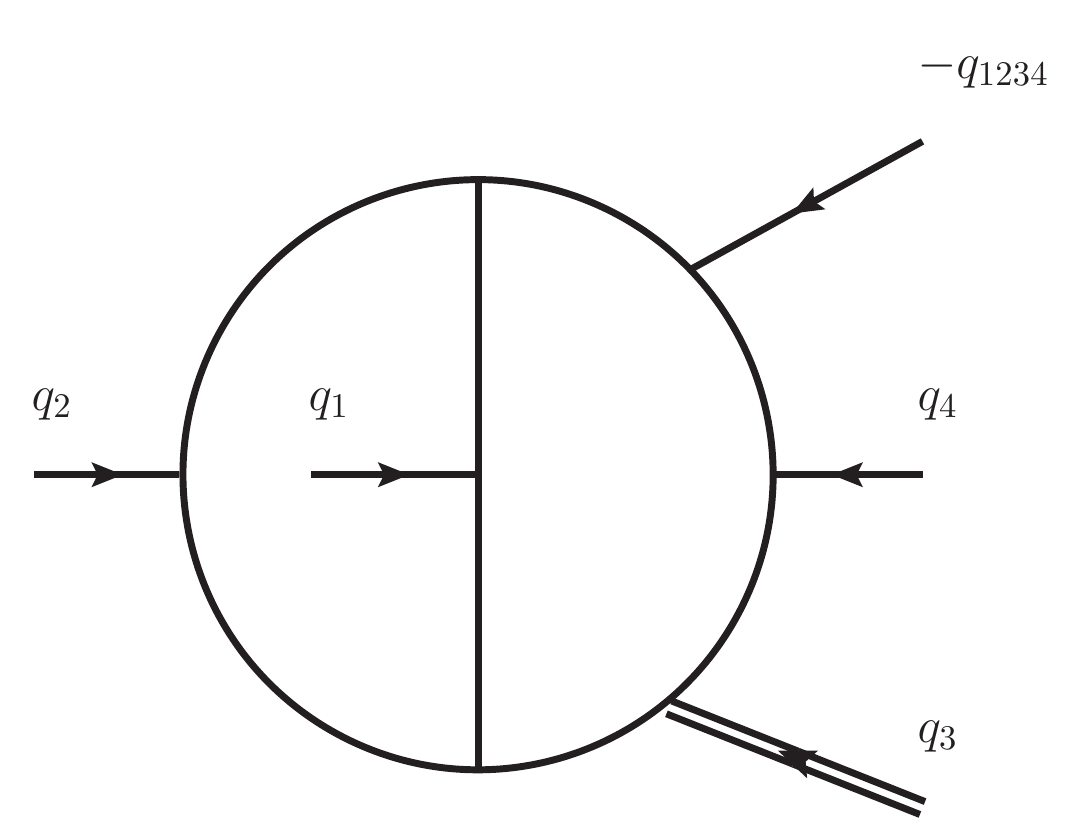} \\
\includegraphics[width=0.3 \linewidth]{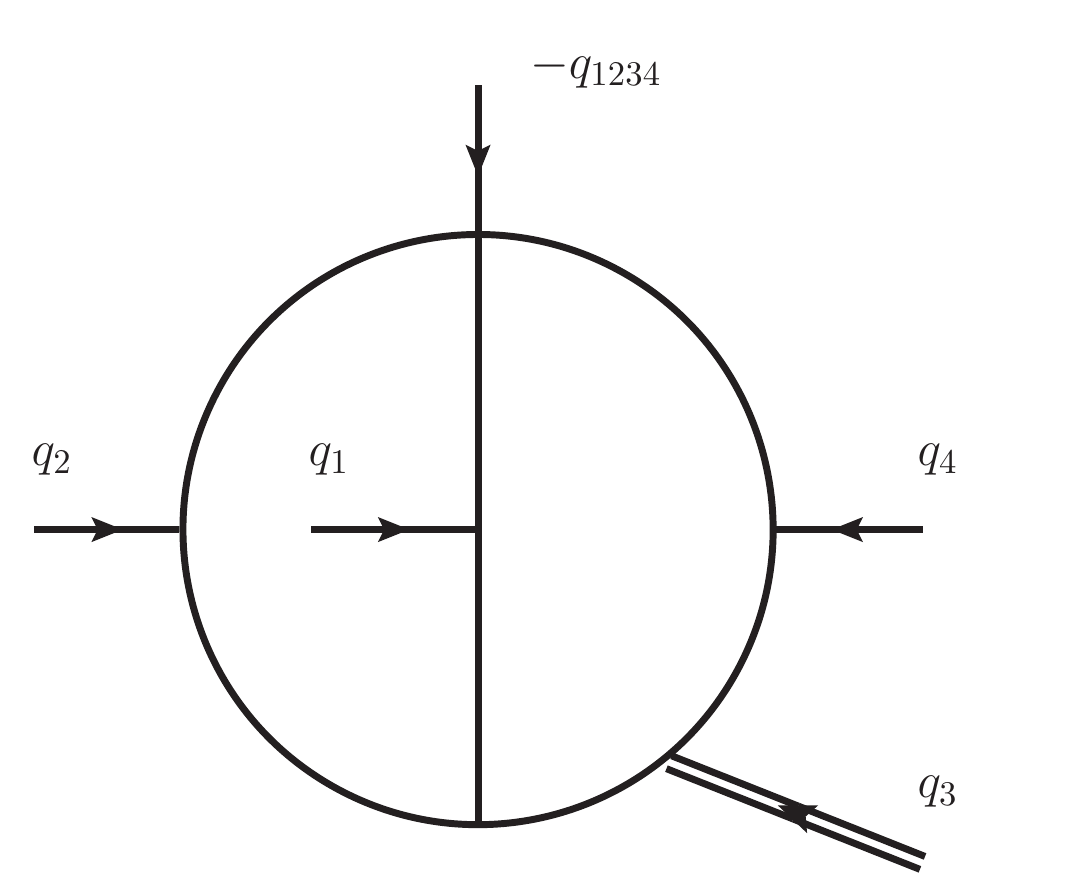} \quad \includegraphics[width=0.3 \linewidth]{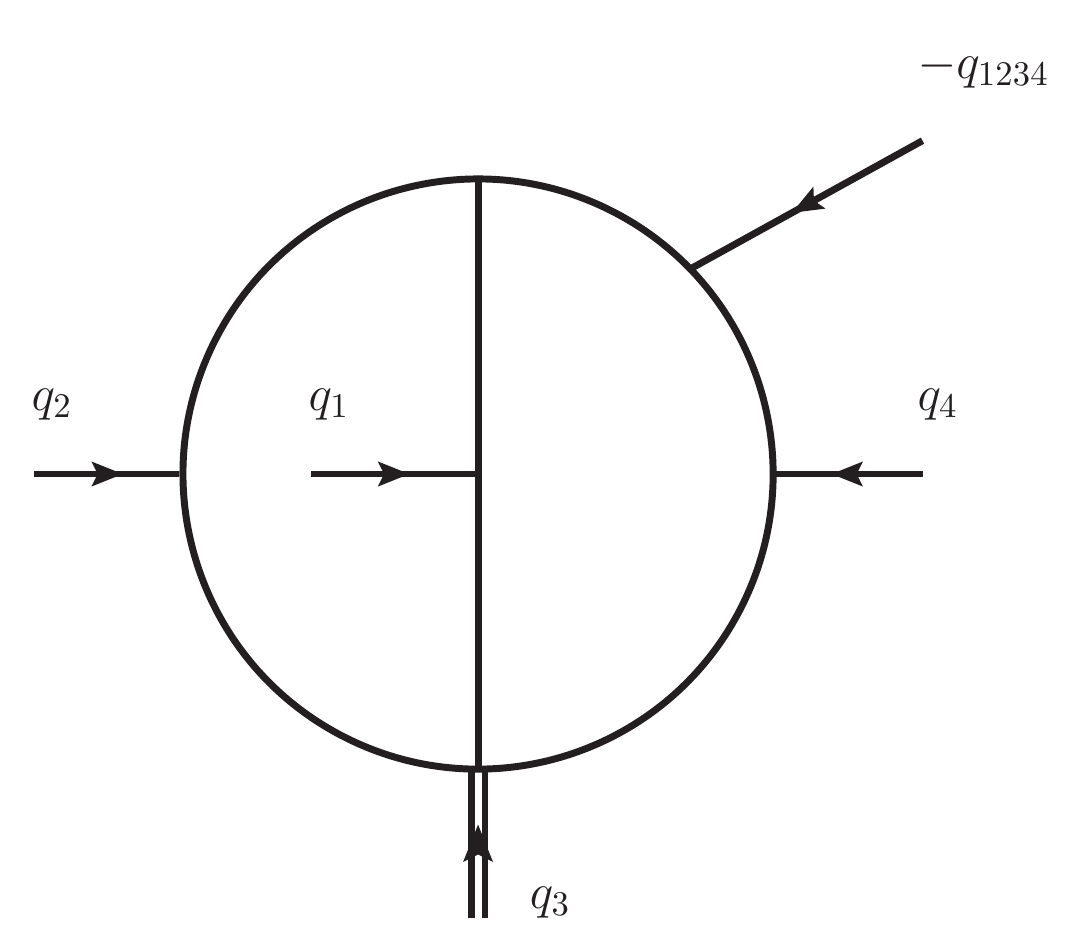} \quad \includegraphics[width=0.3 \linewidth]{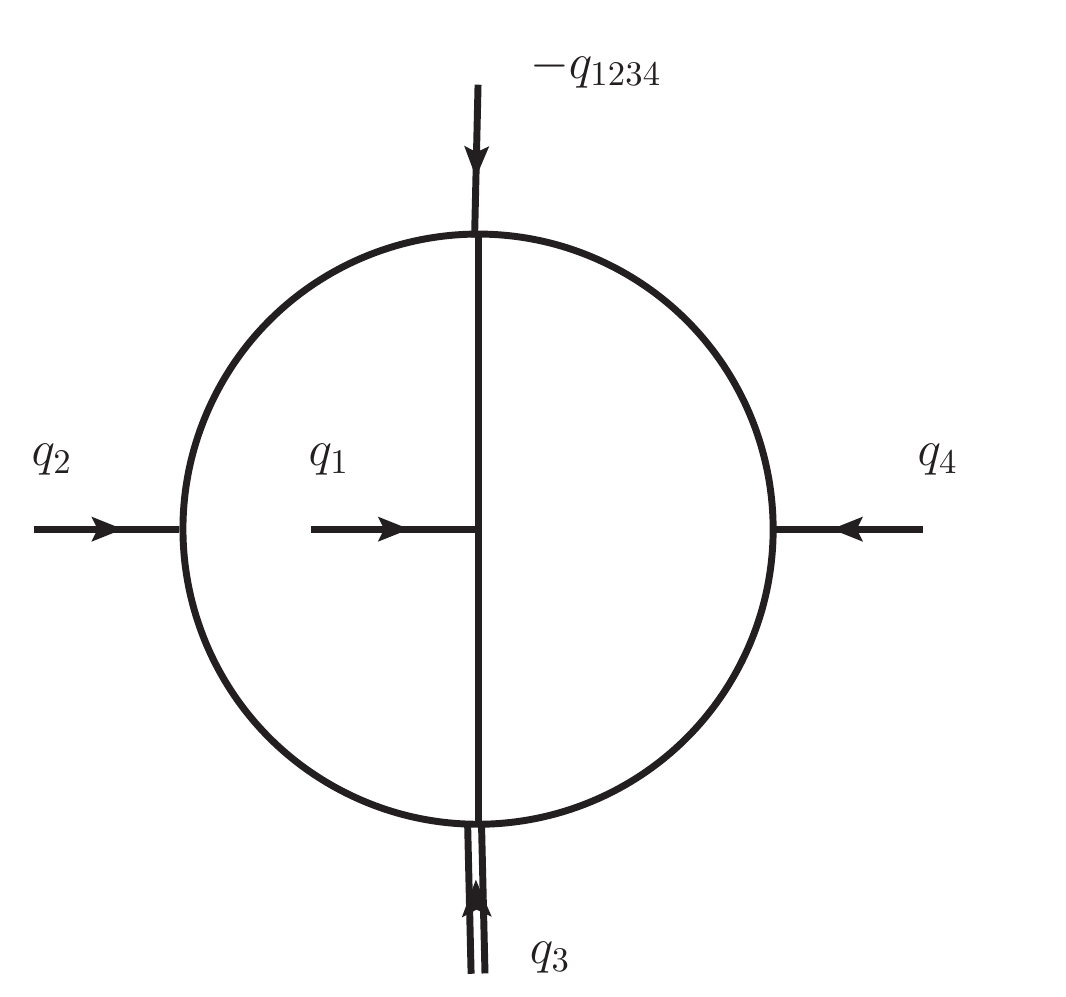} 
\caption{The new two-loop nonplanar five-point diagrams in the family $N_1$. All external momenta are incoming.}
\label{fig::diagnp1}
\end{figure}

In this section we will apply the method discussed in the previous section to a family of planar and nonplanar pentaboxes with one off-shell leg. In particular we will compute the MI of the sectors shown in Figs~\ref{fig::diagp1}, \ref{fig::diagnp1}. The leg with momentum $q_3$ is off-shell, $q_3^2\neq 0$. The integrals shown in Fig.~\ref{fig::diagnp1} are new and have not been computed before.

The planar family of pentabox integrals of which Fig.~\ref{fig::diagp1} is the top sector is defined as
\begin{gather}
G^{P}_{a_1\cdots a_{11}}:=e^{2\gamma_E \epsilon} \int \frac{d^dk_1}{i\pi^{d/2}}\frac{d^dk_2}{i\pi^{d/2}}
\frac{1}{k_1^{2a_1} (k_1 + q_1)^{2a_2} (k_1 + q_{12})^{2a_3} (k_1 + q_{123})^{2a_4}} \nonumber\\
\times \frac{1}{(k_1 + q_{1234})^{2a_5} k_2^{2a_6} (k_2 - q_1)^{2a_7} (k_2 - q_{12})^{2a_8} 
(k_2 - q_{123})^{2a_9} (k_2 - q_{1234})^{2a_{10}} k_{12}^{2a_{11}}}, \label{eq::P1}
\end{gather}
while the nonplanar family of pentabox integrals in Fig.~\ref{fig::diagnp1} is defined as
\begin{gather}
G^{NP}_{a_1\cdots a_{11}}:=e^{2\gamma_E \epsilon} \int \frac{d^dk_1}{i\pi^{d/2}}\frac{d^dk_2}{i\pi^{d/2}}
\frac{1}{(k_2-q_1)^{2a_1} (k_1 + q_1)^{2a_2} (k_1 + q_{12})^{2a_3} k_2^{2a_4}} \nonumber\\
\times \frac{1}{(k_2-q_{12})^{2a_5} (k_2-q_{123})^{2a_6} (k_2 - q_{1234})^{2a_7} k_{12}^{2a_8} 
(k_{12} + q_1)^{2a_9} (k_{12} - q_{1234})^{2a_{10}} (k_{12} - q_3)^{2a_{11}}}. \label{eq::NP1}
\end{gather}
The pentabox integrals in the two families above depend on six variables that we define as,
\begin{equation}
s_{ij}=(q_i+q_j)^2, \quad q_3^2 = m_3^2, \quad ij\in\{12, 23, 34, 45, 51\}. \label{eq::5pinv}
\end{equation}

We note that the MIs inside the planar family $P$ have been computed previously in~\cite{Papadopoulos:2015jft}, but we decided to recompute the three in the top sector using the steps explained in this paper since it serves as a nontrivial check of the method. The top sector, shown in Fig.~\ref{fig::diagp1}, contains the following three integrals,
{\footnotesize
\begin{equation}
\{G^{P}_{11100101111}, G^{P}_{111-10101111}, G^{P}_{111001-11111}\}. \label{eq::PMI}
\end{equation}
}
The nonplanar family $NP$ contains 4 sectors, shown in Fig.~\ref{fig::diagnp1} with ten previously unknown integrals that we choose as follows,
{\footnotesize
\begin{gather}
\{G^{NP}_{-11111111100}, G^{NP}_{-21111111100}, G^{NP}_{-31111111100}, \label{eq::NPMI} \\
G^{NP}_{-11101111100}, G^{NP}_{-21101111100}, G^{NP}_{-111-11111100},
G^{NP}_{-11110111100}, G^{NP}_{-21110111100}, G^{NP}_{-1111-1111100}, 
G^{NP}_{01100111100}\}. \nonumber
\end{gather}
}
All other MI in the family $NP$ lie inside the doublebox families $D,ND$ in Eqs.~\eqref{eq::D}, \eqref{eq::ND} below and have been computed before~\cite{Henn:2014lfa,Caola:2014lpa,Papadopoulos:2014hla,Gehrmann:2015ora}.

We may now use the Internal Reduction method discussed in the previous section to relate the scalar planar pentabox MI to a planar double-box integral as shown below.
\begin{eqnarray}
G^{P}_{11100101111} &=& \quad\vcenter{\hbox{\includegraphics[scale=.35]{figs/two-loop-p5p-t8-eps-converted-to.pdf}}} =\int_{0}^{1} dx_f\quad  \vcenter{\hbox{\includegraphics[scale=.35]{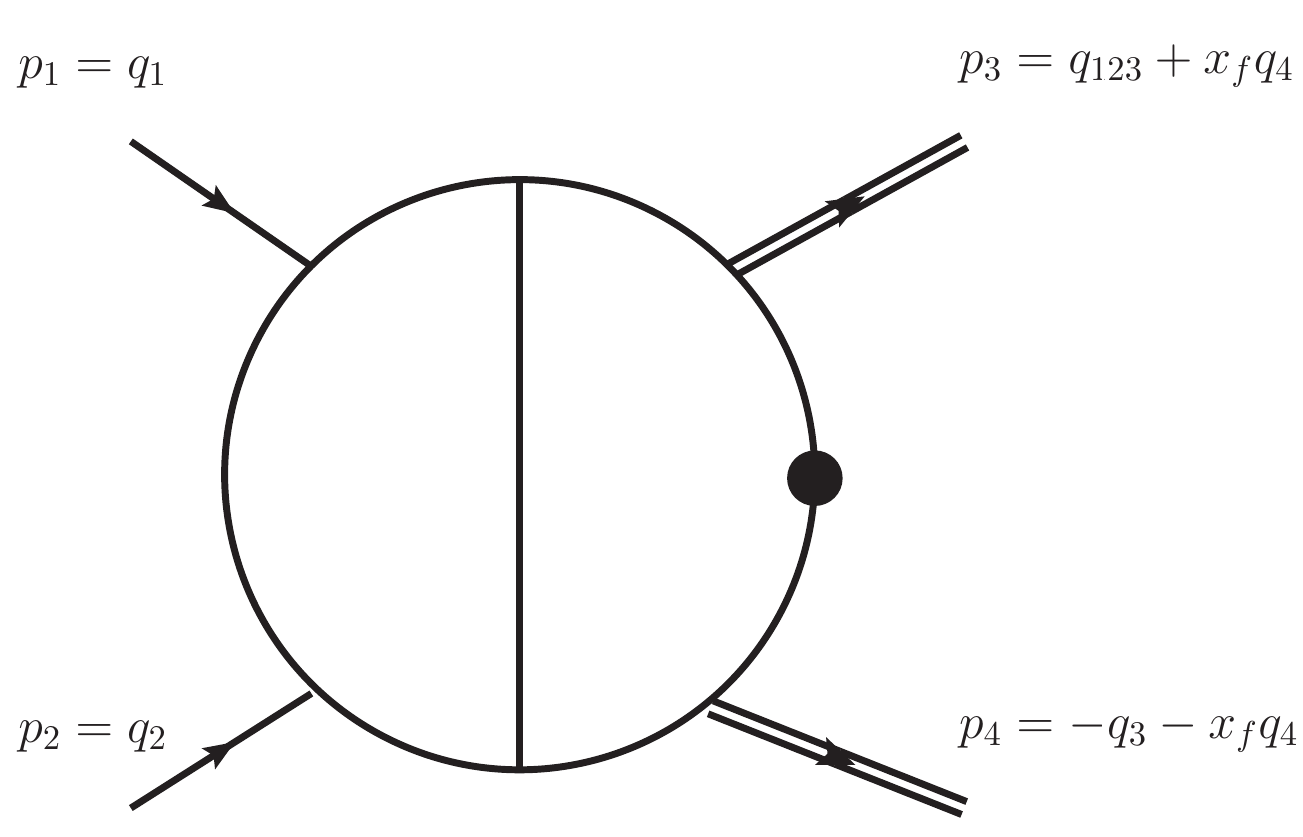}}} \nonumber\\[12pt]
&=&: \int_0^1 dx_f \, G^{D}_{111121100}(x_f). \label{eq::p5to4}
\end{eqnarray}
Similarly, the scalar nonplanar pentabox MI can be related to the non-planar double-box family as shown below.
\begin{eqnarray}
G^{NP}_{01111111100} &=& \quad \vcenter{\hbox{\includegraphics[scale=.35]{figs/two-loop-np5p-t8-eps-converted-to.pdf}}}
=\int_{0}^{1} dx_f\quad  \vcenter{\hbox{\includegraphics[scale=.35]{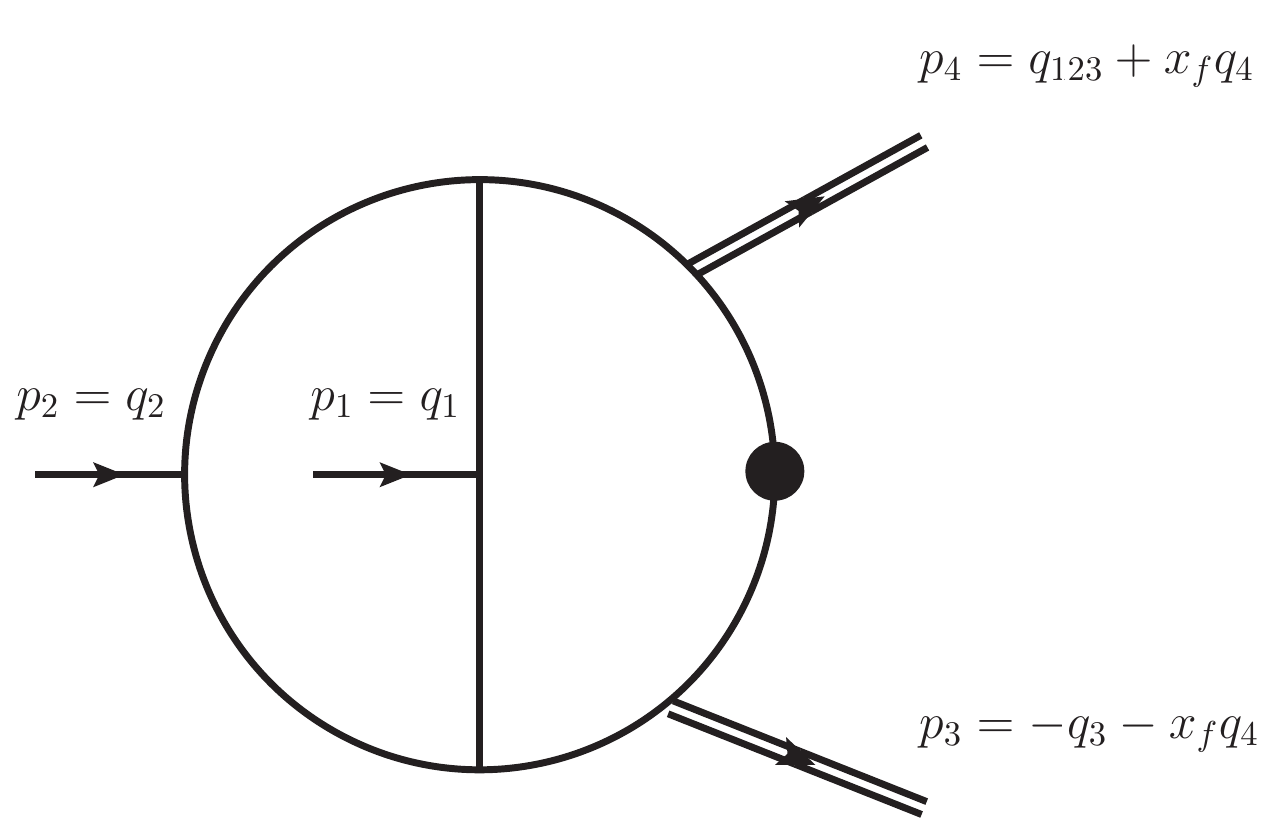}}} \nonumber\\[12pt]
&=&: \int_0^1 dx_f \, G^{ND}_{111121100}(x_f). \label{eq::np5to4}
\end{eqnarray}
The planar and nonplanar double box integral that appear inside the integral on the r.h.s. in Eqs.~\eqref{eq::p5to4} and~\eqref{eq::np5to4} can be IBP reduced to a set of MIs that are known analytically and have been computed in~\cite{Henn:2014lfa,Papadopoulos:2014hla,Gehrmann:2015ora} and~\cite{Caola:2014lpa,Papadopoulos:2014hla,Gehrmann:2015ora} respectively. Their family of integrals is defined as
\begin{gather}
G^{D}_{a_1\cdots a_{9}}:=e^{2\gamma_E \epsilon} \int \frac{d^dk_1}{i\pi^{d/2}}\frac{d^dk_2}{i\pi^{d/2}}
\frac{1}{k_1^{2a_1} (k_1 - p_{34})^{2a_2} k_2^{2a_3} (k_2 - p_{34})^{2a_4}(k_1 - p_3)^{2a_5} (k_1 - k_2)^{2a_6}} \nonumber\\
\times \frac{1}{(k_2 - p_1)^{2a_7}(k_2 - p_3)^{2a_8}(k_1 - p_1)^{2a_9}} \, , \label{eq::D} \\
G^{ND}_{a_1\cdots a_{9}}:=e^{2\gamma_E \epsilon} \int \frac{d^dk_1}{i\pi^{d/2}}\frac{d^dk_2}{i\pi^{d/2}}
\frac{1}{k_1^{2a_1} (k_1 - p_{34})^{2a_2} k_2^{2a_3} (k_1 - k_2 -p_1)^{2a_4}(k_1 - p_4)^{2a_5} (k_1 - k_2)^{2a_6} } \nonumber\\
\times \frac{1}{(k_2 - p_2)^{2a_7}(k_2 - p_4)^{2a_8}(k_1 - p_2)^{2a_9}} \, . \label{eq::ND}
\end{gather}
The four external momenta $p_1, p_2, p_3, p_4$ of the planar doublebox integrals may be expressed in terms of the five-point external momenta of the pentabox and similarly goes for the invariants,
\begin{gather}
p_1 = q_1, \quad p_2 =q_2, \quad p_3=-q_3-x_f q_4, \quad p_4=q_{123}+x_f q_4, \label{eq::4pmom}\\
p_1^2=p_2^2=0, \quad S:=(p_1+p_2)^2=s_{12}, \quad T:=(p_1-p_3)^2 = (1-x_f)s_{23} + s_{51}x_f, \nonumber\\
M_3^2:=p_3^2 = (1-x_f) m_3^2+x_f s_{34}, \quad M_4^2:=p_4^2=(1-x_f)s_{45}. \label{eq::4pinv}
\end{gather}
For the nonplanar doublebox, we flipped $p_3$ and $p_4$ in Eqs.~\eqref{eq::4pmom}, \eqref{eq::4pinv} (cf. Eq.~\eqref{eq::np5to4}).

For the evaluation of double-box integrals we have used a common basis of MI taken from refs.~\cite{Henn:2014lfa,Caola:2014lpa}
and two different parametrizations, namely the one given in~\cite{Henn:2014lfa,Caola:2014lpa} and the second one from~\cite{Papadopoulos:2014hla}.
In both cases the basis elements satisfy the following canonical equation
\begin{equation}
d{g_i} = \epsilon \sum\limits_k^{} {d\log \left( {{\alpha_k}} \right)\sum\limits_j^{} {{\cal M}_{ij}^{\left( k \right)}{g_j}} }
\label{dlog}
\end{equation}
with ${\cal M}^{(k)}$ being constant matrices consisting of rational numbers.   
In the former case the so-called letters, $\alpha_k$, are given by
\begin{align}
& \alpha=\{x, y, z, 1 + x, 1 - y, 1 - z,  1 + x y, z-y,
1 + y(1 + x) - z,  x y + z, \nonumber \\
& 1 + x(1 +  y - z), 1 + x z \},
\end{align}
for the planar family, and
\begin{align}
 & \alpha=\{x, 1 + x, 1 - y, y, 1 + x y, 1 + 
 x (1 + y - z), 1 - z, z - y, 1 + y - z \nonumber \\
& 1 + y + x y - z, z, 
x y + z, 1 + x + x y - x z, 1 + x z\},
\end{align}
for the non-planar family, with $x,y,z$ defined by
\begin{equation}
\frac{S}{M_3^2} = (1+x) ( 1 + x y),\;\;\; \frac{T}{M_3^2} = -xz,\;\;\; \frac{M_4^2}{M_3^2}  = x^2 y,
\end{equation}
whereas in the second case the letters are given by $\alpha_k\equiv\bar{x}-\beta_k$, with 
\begin{align}
    \beta=\left\{0,1,\frac{q}{S_{12}}, \frac{q}{q-S_{23}},1+\frac{S_{23}}{S_{12}},\frac{q-S_{23}}{S_{12}} \right\},
\end{align}
for the planar family, and
\begin{align}
    \beta=\left\{0,1, \frac{q}{S_{12}}, \frac{q}{q-S_{23}},1+\frac{S_{23}}{S_{12}},\frac{q-S_{23}}{S_{12}},
    \frac{q}{S_{12}+S_{23}}\right\},
\end{align}
for the non-planar family, with
$\bar{x},S_{12},S_{23},q$  defined by
\begin{equation}
S = \bar{x}^2 S_{12},\;\;\; T = \bar{x} S_{23}+(1-\bar{x})q,\;\;\; M_3^2=(1-\bar{x})(q-S_{12}\bar{x}),
\;\;\;{M_4^2}  = q .
\end{equation}
In both cases the solution of the equation \eqref{dlog}, is given in terms  
of Goncharov polylogs, ${\cal G}\left( {{w_1}, \ldots ,{w_n};{w_0}} \right)$, with constant rational coefficients. In fact, in the second case Goncharov polylogs have the form ${\cal G}\left( {{\beta _1}, \ldots ,{\beta _n};\bar x} \right)$, 
a distinct characteristic of the simplified differential equations approach~\cite{Papadopoulos:2014lla}, for which Eq.~\eqref{dlog} turns out to be an ordinary differential equation with respect to $\bar{x}$.
The corresponding matrices of rational numbers, ${{\cal M}^{\left( k \right)}}$, for this case, can be found in the ancillary files.


The doublebox integrals in the families $G^{D}$ and $G^{ND}$ are expressed in terms of Goncharov polylogs (GPLs) that can be evaluated numerically 
with the {\tt GiNaC} package~\cite{Vollinga:2004sn}. Though we wrote the relations in Eqs.~\eqref{eq::p5to4}, \eqref{eq::np5to4} only for the scalar integrals $G^{P}_{11001011111}$ and $G^{NP}_{01111111100}$, the same internal reduction technique can be used to relate all planar and nonplanar MIs in Eqs.~\eqref{eq::PMI}, \eqref{eq::NPMI} to double box integrals. These relations are written in appendix~\ref{app:A}.

The integrands on the r.h.s. of Eqs.~\eqref{eq::p5to4},~\eqref{eq::np5to4} contains a pole at $x_f=1$, corresponding to the case when the off-shell leg $p_3$ ($p_4$) of the planar (nonplanar) double box in Eqs.~\eqref{eq::p5to4}, \eqref{eq::np5to4} become on-shell $p_3^2\rightarrow 0$ ($p_4^2\rightarrow 0$). In order to deal with this singularity of the Feynman parameter integral over $x_f$ we need to compute the singular behaviour of the integrands (cf. Eqs.~\eqref{eq::p5to4},~\eqref{eq::np5to4}) at $x_f=1$,
\begin{equation}
\qquad
\begin{gathered}
G^{i}_{111121100}(x_f) = \sum_{a=1,2,4} (1-x_f)^{-1-a\epsilon}\,  G_{111121100,\text{res}}^{i,(a)} + \mathcal{O}((1-x_f)^0), \quad i=D, ND,
\end{gathered}
\label{eq::pxf1}
\end{equation}
where $G_{111121100,\text{res}}^{i,(a)}$ does not depend on $x_f$ but only on the five-point invariants in Eq~\eqref{eq::5pinv} through Eq.~\eqref{eq::4pinv}. We note that the pentabox MI in Eqs.~\eqref{eq::PMI},~\eqref{eq::NPMI} have been chosen such that their corresponding integrands have at most a $(1-x_f)^{n-a\epsilon}$ singularity shown in Eq.~\eqref{eq::pxf1} with $n\geq-1$. The resummed expression for the integrand can be computed by making use of the differential equations of the corresponding doublebox MIs, Eq.~\eqref{dlog}.
We may expand the differential equations around $x_f=1$,
\begin{equation}
\frac{d}{dx_f} \vec{g}_{\text{can}}^{(i)} = \frac{\epsilon {\cal M}^{(i)}}{1-x_f}\cdot \vec{g}_{\text{can}}^{(i)} + \mathcal{O}((1-x_f)^0), \quad i=D, ND. \label{eq::pdiffxf1}
\end{equation}
The eigenvalues of the matrix ${\cal M}^{(i)}$ are $a=1,2,4$ and provide the exponents $-1-a\epsilon$ shown in Eq.~\eqref{eq::pxf1}. Solving Eq.~\eqref{eq::pdiffxf1} is straightforward and results in an expansion of the form given in Eq.~\eqref{eq::pxf1} for each of the canonical integrals. The four-point integrands are then related to these canonical double box integrals through IBP reduction and therefore their resummed expressions around $x_f=1$ are also found. Once the MI have been {\it resummed} in this way, we may regularize the integration in the Feynman parameter around $x_f=1$ in the standard way,
\begin{gather}
\int_0^1 dx_f \, G^{i}_{111121100}(x_f) = \int_0^1 dx_f \, \left(G^{i}_{111121100}(x_f) - \sum_{a=1,2,4} (1-x_f)^{-1-a\epsilon} G_{111121100,\text{res}}^{i,(a)}\right) \nonumber\\
- \sum_{a=1,2,4} \frac{G_{111121100,\text{res}}^{i,(a)}}{a\epsilon}, \quad i=D,ND. \label{eq::pint}
\end{gather}
The remaining integral in Eq.~\eqref{eq::pint} is now integrable in the interval $x_f\in[0,1]$ and may be safely expanded in $\epsilon$. Its evaluation may be performed numerically with any integration routine that links to {\tt GiNaC} to evaluate the GPLs of the four-point integrand $G^i_{111121100}$. 

For the second parametrization of the double-box integrals, we opt to replace the integration over $x_f$ with that over the $\bar{x}$ variable by using the following relation
\begin{equation}
x_f =\frac{s_{12}+\bar{x}\left(s_{45}-s_{12}-m_3^2(1-\bar{x})\right)}
{\bar{x}\left(s_{45}+(s_{34}-m_3^2)(1-\bar{x})\right)}
\label{eq::xtoxf}
\end{equation}
with
${x_f}:\left[ {0,1} \right] \to \bar{x}:\left[ {{\bar{x}_{low}},1} \right]$
and ${\bar{x}_{low}}$ the positive root of $x_f=0$ equation.
The singular limit $x_f\to1$ is given by $\bar{x}\to1$, and the resummed expressions are straightforwardly calculated from the Jordan decomposition of the matrix ${\cal M}^{(k)}$ in Eq.~\eqref{dlog} that corresponds to the letter $\bar{x}-1$. Moreover, this choice produces a square-root free representation for all integrands in Eqs.~\eqref{eq::p5to4} and \eqref{eq::np5to4}. 

In the ancillary files we provide the integrand for the ten nonplanar pentabox MI in Eq.~\eqref{eq::NPMI} in terms of a canonical basis. 
For the canonical basis we have plugged in the two different solutions, described above.
The first one can be used more efficiently in the physical region, where the five-point momenta $q_{1,2}$ are incoming, while the second parametrization in the Euclidean region. 
We have checked numerically that both solutions produce the same results in all regions.
Furthermore, we also provide the resummed expression $G^i_{\text{res}}$ around $x_f=1$ ($\bar{x}=1$) (as in Eq.~\eqref{eq::pxf1}) for the canonical MI of the  
doublebox families $i=D, ND$, plus their expanded versions in $\epsilon$, that is needed in Eq.~\eqref{eq::pint}.
\section{Results and checks} 
\label{sResults}

In this section we discuss the integration over the Feynman parameter $x_f$ ($\Bar{x}$). The simplest way is to evaluate the integral in Eq.~\eqref{eq::pint} numerically in for example {\tt Mathematica}~\cite{Mathematica}. We provide a {\tt Mathematica} notebook in the ancillary files where the reader can load the ingredients of Eq.~\eqref{eq::pint} and evaluate the integral over $x_f$. A very naive integration in {\tt Mathematica} with {\tt NIntegrate} takes about the order of ten minutes to compute all ten integrals on a home laptop to at least 5 digits of precision in a specific point. 

We compare in Table~\ref{tab::res} our results of the numerical integration for the ten new nonplanar pentaboxes with Secdec. The Secdec error is given for the last significant digit in brackets. The error of our analytic result is smaller than the given digits and is left out in the table. For the planar MI we compared our results in this paper with the results computed in~\cite{Papadopoulos:2015jft} and we found excellent agreement, both in the Euclidean and physical region. We have chosen the following Euclidean and physical points,
\begin{eqnarray}
\text{Euclidean:} && \quad s_{12} = -1, \quad s_{23} = -11, \quad s_{34}=-2, \quad s_{45}=-7, \quad s_{51}=-1/2, \quad M_3^2=-31, \nonumber\\
\text{Physical:} && \quad s_{12} = 4, \quad s_{23} = -2.5868, \quad s_{34}=1.55556, \quad s_{45}=0.790123, \quad s_{51}=-2.02801, \nonumber\\
&& M_3^2=0.345679. \nonumber
\end{eqnarray}
We managed to obtain numerical results
in the Euclidean region with Secdec and we find very good agreement as seen from the table. 
We were however not able to obtain numerical results in the physical region with Secdec.

\begin{table}[t!]
\tiny
\begin{center}
\begin{tabular}{|l||c|c||c|}
\hline
\text{} & \multicolumn{2}{c||}{Euclidean} & Physical\\
\hline
\text{} & IntRed & Secdec & IntRed\\
\hline
\multirow{2}{*}{$G^{NP}_{-11111111100}$} &  $\frac{-0.159341}{\epsilon^4}+\frac{-0.697695}{\epsilon^3}+\frac{-1.1696}{\epsilon^2}$ & $\frac{-0.159341(2)}{\epsilon^4}+\frac{-0.697682(6)}{\epsilon^3}+\frac{-1.16946(2)}{\epsilon^2}$ & $\frac{-2.5898}{\epsilon^4}-\frac{6.59609 - 18.6592 i}{\epsilon^3}-\frac{118.267 + 130.474 i}{\epsilon^2}$\\
& $+\frac{2.83058}{\epsilon}+27.6911$ & $+\frac{2.8300(3)}{\epsilon}+27.69(5)$ & $-\frac{193.184 + 1082.88 i}{\epsilon}+884.781 - 2280.76 i$\\
\hline
\multirow{2}{*}{$G^{NP}_{-21111111100}$} &  $\frac{0.0906593}{\epsilon^4}+\frac{0.270172}{\epsilon^3}+\frac{-0.522085}{\epsilon^2}$ & $\frac{0.090659(1)}{\epsilon^4}+\frac{0.270172(2)}{\epsilon^3}+\frac{-0.52212(1)}{\epsilon^2}$ & $\frac{6.12648}{\epsilon^4}+\frac{15.1959 - 38.4938 i}{\epsilon^3}+\frac{306.753 + 281.898 i}{\epsilon^2}$\\
& $+\frac{-7.73903}{\epsilon}-38.4983$ & $+\frac{-7.7388(1)}{\epsilon}-38.4982(7)$ & $+\frac{590.813 + 2423.85 i}{\epsilon}-2296.24 + 4996.11 i$\\
\hline
\multirow{2}{*}{$G^{NP}_{-31111111100}$} &  $\frac{-0.0164835}{\epsilon^4}+\frac{-0.0712228}{\epsilon^3}+\frac{-0.49709}{\epsilon^2}$ & $\frac{-0.016484(1)}{\epsilon^4}+\frac{-0.071221(1)}{\epsilon^3}+\frac{-0.49708(5)}{\epsilon^2}$ & $\frac{-16.6923}{\epsilon^4}-\frac{37.5976 - 85.6334 i}{\epsilon^3}-\frac{756.361 + 527.72 i}{\epsilon^2}$\\
& $+\frac{-1.21945}{\epsilon}-5.20559$ & $+\frac{-1.21941(4)}{\epsilon}-5.2057(7)$ & $-\frac{1519.09 + 5030.47 i}{\epsilon}+6662.12 - 9567.49 i$\\
\hline
\multirow{2}{*}{$G^{NP}_{-11101111100}$} &  $\frac{0.0165809}{\epsilon^3}+\frac{0.151748}{\epsilon^2}$ & $\frac{0.016581(1)}{\epsilon^3}+\frac{0.151748(2)}{\epsilon^2}$ & $\frac{-1.90886}{\epsilon^3}-\frac{14.896 + 123.105 i}{\epsilon^2}$\\
& $+\frac{0.134287}{\epsilon}-0.678448$ & $+\frac{0.134281(4)}{\epsilon}-0.67848(1)$ & $+\frac{66.0738 - 557.165 i}{\epsilon}+229.512 - 775.822 i$\\
\hline
\multirow{2}{*}{$G^{NP}_{-21101111100}$} &  $\frac{0.0103818}{\epsilon^3}+\frac{-0.0651675}{\epsilon^2}$ & $\frac{0.010382(1)}{\epsilon^3}+\frac{-0.065168(1)}{\epsilon^2}$ & $\frac{6.38123 - 5.11369 i}{\epsilon^3}+\frac{91.8762 + 288.091 i}{\epsilon^2}$\\
& $+\frac{0.180387}{\epsilon}-1.44233$ & $+\frac{0.180384(4)}{\epsilon}-1.44233(1)$ & $-\frac{12.5501 - 1510.61 i}{\epsilon}-876.573 + 2486.56 i$\\
\hline
\multirow{2}{*}{$G^{NP}_{-111-11111100}$} &  $\frac{-0.00619909}{\epsilon^3}+\frac{-0.0833771}{\epsilon^2}$ & $\frac{-0.006199(1)}{\epsilon^3}+\frac{-0.083377(1)}{\epsilon^2}$ & $-\frac{1.25421 + 5.11369 i}{\epsilon^3}+\frac{51.2098 - 80.7839 i}{\epsilon^2}$\\
& $+\frac{0.215257}{\epsilon}-0.834736$ & $+\frac{0.215256(4)}{\epsilon}-0.834738(7)$ & $+\frac{239.406 - 126.849 i}{\epsilon}-103.616 + 333.341 i$\\
\hline
\multirow{2}{*}{$G^{NP}_{-11110111100}$} &  $\frac{0.321429}{\epsilon^4}+\frac{1.19884}{\epsilon^3}+\frac{0.905133}{\epsilon^2}$ & $\frac{0.321429(3)}{\epsilon^4}+\frac{1.198837(8)}{\epsilon^3}+\frac{0.90511(2)}{\epsilon^2}$ & $\frac{-3.41912}{\epsilon^4}-\frac{7.51728 - 31.031 i}{\epsilon^3}-\frac{123.246 + 74.0946 i}{\epsilon^2}$\\
& $+\frac{-12.4255}{\epsilon}-78.7869$ & $+\frac{-12.4262(1)}{\epsilon}-78.7897(5)$ & $-\frac{236.233 + 1015.95 i}{\epsilon}+659.761 - 2656.88 i$\\
\hline
\multirow{2}{*}{$G^{NP}_{-21110111100}$} &  $\frac{-0.142857}{\epsilon^4}+\frac{-0.525064}{\epsilon^3}+\frac{-0.797536}{\epsilon^2}$ & $\frac{-0.1428571(1)}{\epsilon^4}+\frac{-0.5250617(5)}{\epsilon^3}+\frac{-0.797531(6)}{\epsilon^2}$ & $\frac{6.16358}{\epsilon^4}+\frac{12.2388 - 52.8607 i}{\epsilon^3}+\frac{309.282 + 212.179 i}{\epsilon^2}$\\
& $+\frac{5.65695}{\epsilon}+33.9655$ & $+\frac{5.661(2)}{\epsilon}+33.979(5)$ & $+\frac{810.965 + 2374.38 i}{\epsilon}-757.756 + 6237.07 i$\\
\hline
\multirow{2}{*}{$G^{NP}_{-1111-1111100}$} &  $\frac{-0.607143}{\epsilon^4}+\frac{-2.31803}{\epsilon^3}+\frac{-3.75067}{\epsilon^2}$ & $\frac{-0.607143(6)}{\epsilon^4}+\frac{-2.31802(2)}{\epsilon^3}+\frac{-3.7501(2)}{\epsilon^2}$ & $\frac{-6.24728}{\epsilon^4}-\frac{16.2843 - 63.311 i}{\epsilon^3}-\frac{138.831 - 31.3273 i}{\epsilon^2}$\\
& $+\frac{17.92}{\epsilon}+126.831$ & $+\frac{17.918(7)}{\epsilon}+126.91(2)$ & $-\frac{98.303 + 947.958 i}{\epsilon}+1460.46 - 2941.73 i$\\
\hline
\multirow{2}{*}{$G^{NP}_{01100111100}$} &  $\frac{0.530639}{\epsilon^2}$ & $\frac{0.530639(5)}{\epsilon^2}$ & $\frac{6.3937 + 40.1341 i}{\epsilon^2}$\\
& $+\frac{1.85246}{\epsilon}+3.7215$ & $+\frac{1.85246(1)}{\epsilon}+3.72150(2)$ & $+\frac{2.38658 + 195.035 i}{\epsilon}-49.2847 + 331.291 i$\\
\hline
\end{tabular}
\end{center}
\caption{Comparison with Secdec. \label{tab::res}}
\end{table}

The numerical integration takes the longest time around $x_f=1$ ($\bar{x}=1$). Therefore one may consider speeding the integration up drastically by splitting the integration interval in $x_f(\bar{x})\in[0,1-\delta]$ and $x_f(\bar{x})\in[1-\delta,1]$, with $\delta\ll 1$. The former evaluates in Mathematica in a manner of seconds, while the latter may be performed analytically by expanding the integrand around $x_f=1$ ($\bar{x}=1$) as explained in the previous section. In this way the numerical integration could in principle be performed in a few seconds on a home laptop. 

One other way of speeding up the evaluation of the integrands is by performing the integrations analytically. For the first parametrization, this may be done by making a variable substitution that rationalizes the square-roots that appear in $x_f$. For example, in the parametrization of ref.~\cite{Henn:2014lfa} the square-root that appears in the integration variable $x_f$ is
\begin{eqnarray}
R_{12} &=& \sqrt{-4M_3^2M_4^2+(M_3^2+M_4^2-S)^2} \\
&=& \sqrt{4 s_{45} (-1 + x_f) (m_3 - m_3 x_f + s_{34} x_f) + (m_3 - s_{12} + s_{45} - m_3 x_f + s_{34} x_f - s_{45} x_f)^2} \nonumber
\end{eqnarray}
and can be rationalized with a new variable $v$ as follows
\begin{gather}
x_f = \frac{s_{12} - s_{12} v + (m_3 + s_{45} (-1 + v)) v}{s_{12} + (m_3 - s_{34} - s_{45}) v}.
\label{eq::varsub}
\end{gather} 
The Goncharov polylogs that one finds have all rational letters in the new variable $v$.
For the second parametrization~\cite{Papadopoulos:2014hla} this is already achieved by the transformation given by Eq.~\eqref{eq::xtoxf}.
By expressing the Goncharov polylogs in the form ${\cal G}(w_1,\ldots,w_n;v)$ (${\cal G}(w_1,\ldots,w_n;\bar{x})$) (e.g. using HyperInt~\cite{Panzer:2014caa}), with weights $w_i$ that do not depend on $v$ ($\bar{x}$), the integral over $v$ ($\bar{x}$) can be performed analytically and the result expressed again in terms of Goncharov polylogs. On the other hand, we expect that analytic results for the canonical basis of the pentabox integral families with one off-shell leg, which is not known yet, will be much more manageable. It is therefore an interesting question how the method of Internal Reduction presented in this paper, can be used to relate directly elements of canonical bases of different families, i.e. those of pentabox and double-box integral families. We
leave these considerations for a future publication.

\section{Conclusions} 
\label{sConc}

   
In this paper we have presented a new method called Internal Reduction. 
By judiciously combining two propagators through a Feynman parameter representation one may reduce multi-scale (e.g. five-point) Feynman graphs to simpler Feynman graphs with less scales (e.g. four-point). Since there is already an expanding database of MI with up to five external legs that are known analytically, we may use this method to recycle these known MIs by relating new unknown multi-leg Feynman integrals to one-fold integrals of these known lower scale MIs.

As a proof of concept, we applied this method to a set of planar and nonplanar two-loop five-point MI with one off-shell leg. The planar family has been computed before and served as a non-trivial check of our implementation. The nonplanar family on the other hand contains ten new MI that are not known. In this paper we derived one-fold integral representations for these ten new MI by using the Internal Reduction method. We compared our results with Secdec in the Euclidean region and found excellent agreement. We have also presented results for these integrals in one of the six physical regions. The extension to the other five physical regions is left for future work. Together with this paper we include the results for the ten new nonplanar integrals in ancillary files. They are expressed in terms of one-fold integrals of Goncharov polylogs that can be evaluated efficiently with Ginac.

For the future, it would be interesting to apply this method to the other topologies of the five-point MI shown in Figs~\ref{fig:param-P} and \ref{fig:param-N}, providing thus the missing ingredients for NNLO QCD computations concerning for instance $V,H+2$ jets production at the LHC. Besides this, one of the possible extensions of the Internal Reduction method is to reduce more internal lines introducing more Feynman parameters. In conclusion, the idea put forward in this paper, namely to relate through integration over Feynman (or possibly externally introduced) parameters more complicated integrals 
to simpler ones, for arbitrary multi-loop MI families,
deserves further investigation. 

\acknowledgments
The research of C.W. was supported in part by the BMBF project No. 05H18WOCA1.
C.G.P. would like to thank CERN Theory Department and TUM for their kind hospitality, during which part of the work has been done.  
\appendix
\crefalias{section}{appsec}
\section{Relating the five-point MI to four-point MI}
\label{app:A}

Here we will relate the five-point planar and nonplanar pentabox MI in Eqs.~\eqref{eq::PMI},~\eqref{eq::NPMI} to planar and nonplanar doublebox integrals in the families $D$, $ND$. First the planar,
\begin{gather}
\{G^{P}_{11100101111}, G^{P}_{111-10101111}, G^{P}_{111001-11111} \} \\
= \int_0^1 dx_f \{G^D_{111121100},G^D_{111121100} +(k_2-q_{123})^2,G^D_{11112110-1}\},
\end{gather}
where in the second component on the right hand side, the doublebox integral contains the numerator $(k_2-q_{123})^2=k_2^2-2 k_2\cdot q_{123}+q_{123}^2$. The $k_2\cdot q_{123}$ part of the integral is dealt with a Pasarino-Veltman reduction and then expressed in terms of integrals lying inside of the doublebox family in Eq.~\eqref{eq::D}. 

For the nonplanar integrals we have the following relations,
{\footnotesize
\begin{gather}
\{G^{NP}_{-11111111100}, G^{NP}_{-21111111100}, G^{NP}_{-31111111100}, \\
G^{NP}_{-11101111100}, G^{NP}_{-21101111100}, G^{NP}_{-111-11111100},
G^{NP}_{-11110111100}, G^{NP}_{-21110111100}, G^{NP}_{-1111-1111100}, 
G^{NP}_{01100111100}\}  \nonumber\\
= \int_0^1 dx_f \{G^{ND}_{11112110-1},G^{ND}_{11112110-2},G^{ND}_{11112110-3}, \\
G^{ND}_{10112110-1},G^{ND}_{10112110-2},G^{ND}_{1-1112110-1}, G^{ND}_{01112110-1},G^{ND}_{01112110-2},G^{ND}_{-11112110-1}, G^{ND}_{001121100}\}.
\end{gather}
}

\def\bibsection{\section*{\refname}}

\bibliographystyle{JHEP}
\bibliography{biblio}

\end{document}